\documentclass[11pt,a4paper]{article}

\usepackage{jheppub} 

\usepackage{graphicx}

\DeclareMathOperator{\Tr}{Tr}
\newcommand{\order}[1]{\mathrm{O}\left( #1 \right)}

\newcommand{\mre}{\mathrm{e}}
\newcommand{\mrd}{\mathrm{d}}

\newcommand{\IDR}{\overline{I}}

\newcommand{\mcWov}{\overline{\mathcal{W}}}
\newcommand{\Wov}{\overline{W}}
\newcommand{\FV}{\mathrm{FV}}

\newcommand{\mcVD}{\mathcal{V}_D}

\newcommand{\gbar}{\overline{g}}
\newcommand{\ms}{\mathrm{{MS\kern-0.14em}\kern0.14em}}
\newcommand{\msbar}{\mathrm{\overline{MS\kern-0.14em}\kern0.14em}}

\newcommand{\J}{\mathrm{J}}

\newcommand{\Lt}{L_t}
\newcommand{\ellhat}{\hat{\ell}}

\newcommand{\FT}{\mathrm{FT}}
\newcommand{\IFT}{\mathrm{IFT}}

\newcommand{\Sp}{S_{\mathrm{p}}}
\newcommand{\Km}{K_{\mathrm{m}}}
\newcommand{\Kpp}{K_{\mathrm{pp}}}

\newcommand{\RE}{\mathrm{Re}}
\newcommand{\IM}{\mathrm{Im}}

\title{\boldmath Casimir squared correction to the standard
rotator Hamiltonian for the O($n$) sigma-model in the delta-regime}

\author[a]{F.\ Niedermayer} \author[b]{and P.\ Weisz}

\affiliation[a]{Albert Einstein Center for Fundamental Physics, \\
  Institute for Theoretical Physics, University of Bern, 
  Switzerland} 
\affiliation[b]{Max-Planck-Institut f\"ur Physik, 80805
  Munich, Germany}

\emailAdd{niedermayer@itp.unibe.ch} \emailAdd{pew@mpp.mpg.de}

\abstract{In a previous paper we found that the
  isospin susceptibility of the O($n$) sigma-model 
  calculated in the standard rotator approximation differs 
  from the next-to-next to leading order chiral perturbation theory result 
  in terms vanishing like $1/\ell\,,$ for $\ell=L_t/L\to\infty$ and further
  showed that this deviation could be described by a correction to the
  rotator spectrum proportional to the square of the quadratic 
  Casimir invariant. Here we confront this expectation with analytic 
  nonperturbative results on the spectrum in 2 dimensions, 
  by Balog and Heged\"us for $n=3,4$ and by Gromov, Kazakov and Vieira 
  for $n=4$.
  We also consider the case of 3 dimensions.}
 
\subheader{\hfill MPP-2017-214 }

\begin{document}

\maketitle

\section{Introduction}
\label{Intro}

In the pioneering paper \cite{Leutwyler:1987ak} Leutwyler 
showed that to lowest order in chiral perturbation theory ($\chi$PT)
the low energy dynamics of QCD in the $\delta-$regime
is described by a quantum rotator for the spatially
constant Goldstone modes. We recall that for a system in a
periodic spatial box of sides $L$ the $\delta-$regime
is where the time extent $\Lt \gg L$
and $m_\pi L$ is small (i.e. small or zero quark mass) 
whereas $F_\pi L$, ($F_\pi$ the pion decay constant) is large.

There are other important physical systems, in particular in condensed
matter physics where anti-ferromagnetic layers are described by 
the O(3) sigma-model for $d=3$ \cite{Hasenfratz:1993vf}, 
where the order parameter of the
spontaneous symmetry breaking is an O(3) vector. 
In the analogous perturbative regime these systems and also the non-linear
sigma models in $d=2$ are described by a quantum rotator to leading order. 

In all such systems the lowest energy momentum zero
states of isospin $I$ have to leading order $\chi$PT energies of the form 
\begin{equation}\label{E_Casimir}
  E_I \propto\mathcal{C}_{n;I}\,,
\end{equation}
where 
\begin{equation}\
\mathcal{C}_{n;I}=I(I+n-2)\,,
\end{equation}
is the eigenvalue of the quadratic Casimir for isospin $I$.

At 1-loop level it turns out that the Casimir scaling \eqref{E_Casimir}
still holds, but it is of course expected that at some higher order 
the standard rotator spectrum will be modified. 
The standard rotator describes a system where the length of the total
magnetization on a time-slice does not change in time. This is obviously
not true in the full effective model given by $\chi$PT.
To our knowledge, the actual deviation from the Casimir scaling 
was first observed by Balog and Heged\"us 
\cite{Balog:2009ze} in their computation of the spectrum of the $d=2$
O(3) non-linear sigma model in a small periodic box (circle) using 
the thermodynamic Bethe ansatz (TBA).

Of course, a deviation from the standard rotator spectrum
could be established by explicit perturbative computations.
However, in our previous paper \cite{Niedermayer:2017uyr}
we pointed out that by comparing the already obtained 
NNLO results for the isospin susceptibility calculated in
$\chi$PT at large $\ell\equiv L_t/L$
with that computed using
the standard rotator one can establish, under reasonable assumptions,
that the leading correction to the rotator Hamiltonian
occurs at 3-loops and is proportional to the square of 
the Casimir operator with a proportionality constant determined 
by the NNLO LEC's of $\chi$PT. 

In Appendix~\ref{spectrumdeq1} we illustrate that
such corrections are expected by recalling the spectrum
of the O(3) rotator in $d=1$ with lattice regularization.
Note that the appearance of terms proportional to 
$\mathcal{C}_I^2$ is not a lattice artifact.
This simple example serves only to show that distorting
the Lagrangian of the standard 1d rotator by an O(3) invariant
perturbation leads naturally to a spectrum which
(at a higher order) contains such term.
 
After reviewing some preliminary results in Sect.~\ref{oldresults},
in Sect.~\ref{cased2} we test our claim above in the $d=2$
O(3) and O(4) non-linear sigma models in a periodic box. 
For O(3) we find excellent agreement of our prediction 
with the analytic computations of the lowest isospin 1 and 2 
energies for O(3) by Balog and Heged\"us \cite{Balog:2009ze}. 

For O(4) Balog and Heged\"us \cite{Balog:2003yr} 
computed only the ground state energy; 
Gromov, Kazakov and Vieira \cite{Gromov:2008gj} computed 
also higher state energies but the results at small volumes
presented there are not sufficiently precise for our purposes.
We thus generated more data; our methods used to solve
the TBA equations are described in Appendix~\ref{nu012}.
Again there is good agreement with our prediction. 

The derivation of our result depends on the validity of
a (plausible) assumption; but also there is, to our knowledge,
no rigorous derivation of the TBA equations (or even the S-matrix)
from first principles starting with the 2d O($n$) QFT%
\footnote{except for $n=4$ which is also a principal chiral model}.
Hence the agreement of the results provides
extra evidence for the validity of both scenarios.

Finally in Sect.~\ref{cased3} compute the effect for the $d=3$ O($n$) model,
but we have not yet found data for which a comparison can be made.

\section{The isospin susceptibility}
\label{oldresults}

In this section we recall some results on the isospin susceptibility.
The Hamiltonian of the O($n$) standard quantum rotator with a chemical 
potential coupled to the generator $\hat{L}_{12}$ of rotations in 
the $12$--plane is
\begin{equation}
  H_0(h) = \frac{\hat{L}^2}{2\Theta} - h \hat{L}_{12} \,,
  \label{H0h} 
\end{equation}
where $\Theta$ is the moment of inertia. In $d=4$ dimensions
to lowest order $\chi$PT one has $\Theta\simeq F^2 L^3$.
The isospin susceptibility is defined as the second derivative 
of the free energy wrt $h$:
\begin{equation}
  \chi= \left.\frac{1}{\Lt L^{d-1}}
  \frac{\partial^2}{\partial h^2}\ln Z(h;\Theta)\right|_{h=0}\,, 
  \quad\quad Z(h;\Theta)=\Tr \exp\{-H_0(h)L_t\}\,.
\end{equation}
In ref.~\cite{Niedermayer:2017uyr} we showed that for small $u=\Lt/(2\Theta)$ 
the isospin susceptibility computed from the standard rotator,
which we call $\chi_{\text{rot}}$ is given by
\begin{equation}
\chi_{\text{rot}}= \frac{\Lt}{nL^{d-1}u}\left[1-\frac13(n-2)u
+\frac{1}{45}(n-2)(n-4)u^2 + \ldots \right]\,.
\label{chi_rot0}
\end{equation}

On the other hand in a previous paper \cite{Niedermayer:2016yll},
we computed the isospin susceptibility in an asymmetric $\Lt\times L^{d-1}$
box (with periodic boundary conditions) and the mass gap, in this
case the lowest energy in the isospin 1 channel, to
NNLO (next-to next-to leading order) $\chi$PT. For the susceptibility
we recall the results in eqs.~(3.54)--(3.57) of \cite{Niedermayer:2016yll}
with dimensional regularization:
\begin{equation}
  \chi = \frac{2}{n g_0^2} \left( 1 + g_0^2 R_1
    + g_0^4 R_2 + \ldots \right)\,,
  \label{chiDR}
\end{equation}
with
\begin{equation}
  R_1 = -2(n-2)\IDR_{21;D}\,,
  \label{R1f}
\end{equation}
and
\begin{equation}
  R_2=R_2^{(a)}+R_2^{(b)}\,,
\end{equation}
with
\begin{equation}
  R_2^{(a)} = 2(n-2)\left\{ -\Wov
    +2\IDR_{21;D}\left[\IDR_{10;D}-\IDR_{21;D}\right]
    +\frac{2(n-2)}{V_D}\IDR_{31;D}\right\}\,,
  \label{R2af}
\end{equation}
and $R_2^{(b)}$ involves the 4--derivative couplings which we shall
not include in this paper. The expressions above require
some explanation. Firstly   
$g_0$ is the bare coupling for $d=2$; $g_0^{-2}=\rho_s$,
the spin stiffness for $d=3$; and $g_0^{-2}=F^2$ for $d=4$ 
where $F$ is the pion decay constant in the chiral limit.
$V_D$ is the volume $V_D=\Lt L^{D-1}\ellhat^q$ where with DR we have
added $q=D-d$ extra dimensions with extent $L\ellhat\,;$ in this paper
we will usually set $\ellhat=1$ unless stated otherwise.
(Note that the renormalized quantities do not depend on $\ellhat$.)
$\IDR_{nm;D}$ are 1-loop dimensionally regularized sums over momenta
(cf eq.~(3.44) of \cite{Niedermayer:2016yll})
\footnote{Note in \cite{Niedermayer:2016yll} we dropped the label
$D$ and wrote only $\IDR_{nm}.$}. Finally $\Wov$ in  
\eqref{R2af} is an integral associated with a two-loop vacuum massless 
sunset diagram, which is discussed in Appendix~\ref{sunset_ell}.

The lowest energy state above the vacuum (the mass gap) is
given by $E_1=m_1$ in eqs.~(5.9)-(5.11) of \cite{Niedermayer:2016yll}:
\begin{equation}\label{energygap}
  E_1=\frac{n_1g_0^2}{2\overline{V}_D}\left[1+
    g_0^2\triangle^{(2)}+g_0^4\triangle^{(3)}+\dots\right]\,,
\end{equation}
(here $\overline{V}_D=L^{D-1}\ellhat^q$), with
\begin{align}
  \triangle^{(2)}&=(n-2)R(0)\,,
  \\
  \triangle^{(3)}&=(n-2)\left[
    2W+\frac{3}{4\overline{V}_D}\IDR_{10:D-1}+(n-3)R(0)^2\right]
  +\text{4-derivative\,\,terms}\,.
\end{align}
Here $R(z)$ is the propagator for an infinitely long strip without the slow
modes and $W$ is a 2-loop sunset integral discussed in 
Appendix~\ref{sunset_ell}.

For the simple rotator \eqref{H0h} at zero isospin chemical potential,
the energy gap is given by
\begin{equation}\label{E1toTheta}
E_1(L)=\frac{(n-1)}{2\Theta(L)}\,.
\end{equation}
By inserting the expression for $\Theta$ using \eqref{energygap}  
into \eqref{chi_rot0},
we obtain the susceptibility for small $u$ as a function of $F,L,\ell$.
This can then be compared to the direct $\chi$PT computation 
\eqref{chiDR} in the $\epsilon$--regime for $\ell \gg 1$.
The comparison requires knowledge of the large $\ell$--behavior
of shape functions and the sunset integral appearing in the latter,
which are discussed in Appendix~\ref{sunset_ell} 
of ref.~\cite{Niedermayer:2017uyr} for $d=4$.

The two results in NNLO differ by $\propto 1/\ell$ terms for $\ell\gg 1$,
(plus terms vanishing exponentially with $\ell$):
\begin{equation} \label{dchi}
    \frac{\chi - \chi_{\text{rot}}}{\chi} =
    \frac{1}{F^4 L^4}\left( \frac{\Delta_2}{\ell} + \ldots\right)
    +\order{\frac{1}{F^{6} L^6}}\,,
\end{equation}
for $d=4$ and similarly for $d=2,3$.

In ref.~\cite{Niedermayer:2017uyr} we showed that the deviation above 
can be accounted for if the spectrum to the order we are considering 
is given by a modified rotator with eigenvalues of the form  
\begin{equation}  \label{ml}
  E_I(L)=\frac{1}{2\Theta(L)}\mathcal{C}_{n;I}
  +\frac{\Phi(L)}{L}\mathcal{C}_{n;I}^2\,.
\end{equation}
$\Theta(L)$ appearing here contains a higher order correction
with respect to the one in \eqref{E1toTheta}, denoted below
by $\Theta_\text{old}(L)$.
\begin{equation}  \label{Theta_new}
  \frac{1}{\Theta(L)} = \frac{1}{\Theta_\text{old}(L)}
  -\frac{2(n-1)}{L} \Phi(L) \,.
\end{equation}
The leading order of the coefficient $\Phi(L)$ in \eqref{ml}
is directly related to $\Delta_2$ in \eqref{dchi}.
In \cite{Niedermayer:2017uyr} we considered only the case $d=4\,;$ 
in the next sections we consider the lower dimensions $d=2,3$.

It is interesting to note that the correction discussed here
shows up at NNLO in the isospin susceptibility $\chi$,
and in the NNNLO (next order) in the spectrum of the effective 
rotator. The reason is that in the (standard) rotator approximation
the Boltzmann weight is $\sim \exp(-g^2 C_I L_t/L) =
\exp(-g^2 \ell C_I)$ where $g^2=g^2(1/L)$. 
The typical values of the isospin in the partition function are then 
given by $C_I \approx 1/(g^2 \ell)$,
hence the extra factor  $\exp( -C_I^2 \Phi(L) L_t/L)$ in the 
Boltzmann weight gives a correction  
$\delta\chi/\chi \sim -\Phi(L)/(g^4\ell)$ for the susceptibility.
With $\Phi(L)=\Phi_3 g^8 + \order{g^{10}}$ 
(cf. \cite{Niedermayer:2017uyr}), 
the leading $\propto C_I^2$ term in the effective Hamiltonian 
which is of NNNLO, results in an NNLO,  $\propto g^4/\ell$ term 
in $\delta\chi/\chi$.

\section{\boldmath Delta regime in $d=2$}
\label{cased2}

The susceptibility computed in $\chi$PT is for $d=2$ given by
(cf eq.~(3.67) of \cite{Niedermayer:2016yll}) 
\begin{equation}
  \chi=\frac{2}{n\gbar_\ms^2(1/L)}
   -\frac{(n-2)}{2\pi n}\gamma_2^{(2)}(\ell)
   -\frac{(n-2)}{2\pi^2 n}r_2(\ell)\gbar_\ms^2(1/L)
    +\ldots
  \label{chi_d2}
\end{equation}
where $\gbar_\ms(q)$ is the minimal subtraction (MS) scheme running
coupling at momentum scale $q$.
The function $r_2(\ell)$ appearing above
is given by (cf eq.~(3.68) of \cite{Niedermayer:2016yll})   
\begin{equation}
  r_2(\ell)=\overline{w}(\ell)-2\kappa_{10}(\ell)-\frac12\gamma_2^{(2)}(\ell)
  \left(\alpha_1^{(2)}(\ell)-\frac{1}{\ell}-\frac12\gamma_2^{(2)}(\ell)\right)
  -\frac{(n-2)}{4\ell}\left(\gamma_3^{(2)}(\ell)+1\right)\,,
  \label{r2}
\end{equation}
where $\overline{w}(\ell)$ appears in the expansion of the sunset
diagram defined (see \eqref{Psi_d2}).

For large $\ell$
(cf (B.20), (B.25), (B.28)-(B.30) of \cite{Niedermayer:2017uyr}): 
\begin{align}
\alpha_1^{(2)}(\ell) &\simeq \alpha_{1/2}^{(1)}(1) +\frac{\pi}{3}\ell
    -2 + \frac{1}{\ell}\,,  
\\ \label{gamma22ell}  
\gamma_2^{(2)}(\ell) &\simeq \alpha_{1/2}^{(1)}(1) -2 + \frac{2\pi}{3}\ell\,,
\\
\gamma_3^{(2)}(\ell)&\simeq\alpha_{3/2}^{(1)}(1)-\frac23
   +\frac{4\pi^2}{45}\ell^3\,.
\end{align}

The susceptibility computed from the simple rotator is given by
(cf eq.~(2.4) of \cite{Niedermayer:2017uyr}):
\begin{equation}\label{chirot_d2}
  \chi_{\mathrm{rot}} = \frac{2}{n \gbar_{\FV}^2(L) }
  -\frac{(n-2)}{3n}\ell  + \frac{(n-2)(n-4)}{90 n}\ell^2 \gbar_{\FV}^2(L)
  +\ldots\,, \quad (d=2)\,,
\end{equation}
where $\gbar_{\FV}$ is the LWW running coupling \cite{LWW} 
defined through the finite volume mass gap: 
 \begin{equation}\label{FVcoupling}
   \gbar_{\FV}^2(L)\equiv \frac{2}{n-1} L E_1(L)\,.
 \end{equation}
Its expansion in terms of the running coupling in the MS scheme
of DR is given by (cf eq.~(2.2) in \cite{LWW}):
 \begin{equation}\label{gFVgMS}
   \gbar_{\FV}^2(L) = \gbar_{\ms}^2(1/L) + c_1 \gbar_{\ms}^4(1/L)
   + c_2 \gbar_{\ms}^6(1/L) +\cdots
 \end{equation}
The first two coefficients were computed in ref.~\cite{LWW}
 \begin{align}
   c_1 &= -\frac{(n-2)}{4\pi}Z\,,
\\
   c_2 &= c_1^2 + \frac{c_1}{2\pi} + \frac{3(n-2)}{16\pi^2}\,,
 \end{align}
where
\begin{equation} \label{Zdef}
Z\equiv \ln(4\pi)-\gamma\,,\quad\quad(\gamma=-\Gamma'(1))\,.
\end{equation}

The 3-loop coefficient $c_3$ can be obtained by combining
Shin's 3-loop computation \cite{Shin1} of the 
finite volume mass gap using lattice regularization,
with the computation of the 3-loop coefficient of the lattice beta--function 
by Caracciolo and Pelissetto \cite{CP}.
\footnote{Note that the original article ref.~\cite{CP} had a few misprints,
some of which were noticed by Shin \cite{Shin2}; 
the final correct result was first presented in ref.~\cite{ACPP}.}
The result is given by
\begin{equation}
c_3-2c_1c_2+c_1^3=\frac{(n-2)}{(4\pi)^3}\left[S_2(n-2)^2+S_1(n-2)+S_0\right]\,,
\end{equation}
with 
\begin{align} \label{S0}
  S_0&=64\pi^3\left(s_1+2s_2+3s_3-0.02730084372483\right)\,,
  \\ \label{S1}
  S_1&=64\pi^3\left(s_2+3s_3-0.00248186195281\right)\,,
  \\ \label{S2}
  S_2&=64\pi^3\left(s_3+0.00004347976248\right)\,,
\end{align}
where we have used Veretin's precise numerical values \cite{Veretin} 
of the lattice integrals appearing in the formulae,
and the $s_i$ are Shin's constants (appearing in eq.~(2.25) of \cite{Shin1})
for which he quotes the following numerical values 
(in eqs.~(2.28)--(2.30) of \cite{Shin1}):
\begin{align} \label{s1}
  s_1&=\phantom{-}0.02903(1)\,,
  \\ \label{s2}
  s_2&=\phantom{-}0.000756(1)\,,
  \\ \label{s3}
  s_3&=-0.000649(1)\,.
\end{align}

Next we insert the expansion \eqref{gFVgMS} in \eqref{chirot_d2} 
and compare the resulting expression with the perturbative result given 
in \eqref{chi_d2}.  
First we see that the $\gbar_{\ms}^0(1/L)$ terms match for large $\ell$ if
\begin{equation}
  \frac{(n-2)}{2\pi}\gamma_2^{(2)}(\ell)\simeq
  2c_1+\frac{(n-2)}{3}\ell\,.
\end{equation} 
This is satisfied since (using \eqref{gamma22ell})
\begin{equation}
  \alpha_{1/2}^{(1)}(1)=2-Z\,,
\end{equation}
which is a special case of the general relation
\begin{equation}
  \alpha_s^{(1)}(1)= \alpha_{1/2-s}^{(1)}(1)=
  \frac{1}{s(1-2s)}+2\pi^{-s}\Gamma(s)\zeta(2s)\,.
\end{equation}

Matching at order $\gbar_{\ms}^2(1/L)$ requires
\footnote{Note the terms proportional to $(n-2)$ match.}
\begin{equation}
\overline{w}(\ell)-2\kappa_{10}(\ell)\simeq
\frac{2\pi^2}{45}\ell^2-\frac{\pi}{6}Z\ell
+\frac34-\frac{Z}{2}+\frac{Z^2}{4}+\order{1/\ell}\,.
\end{equation}
Using eq.~(\ref{R2T2reld2}) and the fact that $\kappa_{10}(\infty)$
is finite consistency requires
\begin{equation} 
  p_1 = 2\kappa_{10}(\infty)
  -\frac{\pi^2}{4} +\frac12 +\frac14 (Z-1)^2\,.
\end{equation}
Using 
\begin{equation}
\kappa_{10}(\infty)=-0.024787427871346432238\,,
\end{equation}
consistency is obtained if
\begin{equation}
  p_1= -1.789538253208505789488\,.
\end{equation}

We arrive at the result
\begin{equation}\label{chidiffPT_d2}
  \frac{\chi-\chi_{\mathrm{rot}}}{\chi}
  =  \frac{(n-2)(n+1)}{16\pi^3 \ell}
  \zeta(3)\, \gbar_{\ms}^4(1/L)+\dots
\end{equation}

On the other hand, from our considerations of the modified rotator
(cf eq.~(11.5) of \cite{Niedermayer:2017uyr}) we would expect 
\begin{equation}\label{chidiffrot_d2}
  \frac{\chi-\chi_{\mathrm{rot}}}{\chi}
  = -\frac{4}{\ell}\Phi_3(n+1)\gbar_{\ms}^4(1/L)+\dots
\end{equation}
where $\Phi_3$ is the leading coefficient in the perturbative expansion
of $\Phi(L)$, assuming the expansion starting at order $\gbar_{\ms}^8$:
\begin{equation}
  \Phi(L)=\sum_{r=3}\Phi_r\gbar_{\ms}^{2r+2}(1/L)\,.
\end{equation}
Comparing \eqref{chidiffPT_d2} with \eqref{chidiffrot_d2} determines 
\begin{equation}
  \Phi_3=-\frac{(n-2)}{(2\pi)^3}f_3
\end{equation}
with
\begin{equation}\label{f3}
  f_3= \frac18 \zeta(3) =0.15025711290\,.
\end{equation}

As a consequence, the low-lying spectrum to order $\gbar_\ms^8$ is given by
\begin{equation}
  \begin{aligned}
    LE_I(L)&=\frac12\mathcal{C}_{n;I}\gbar_\ms^2(1/L)
    \Big\{ 1+c_1\gbar_\ms^2(1/L)+c_2\gbar_\ms^4(1/L)
    \\
    & +\overline{c}_3 \gbar_\ms^6(1/L) + \ldots \Big\}
    + \mathcal{C}_{n;I}^2 \Phi_3\gbar_\ms^8(1/L) 
    +\dots
  \end{aligned}
\end{equation}
where
\begin{equation}
  \overline{c}_3 = c_3 - 2(n-1) \Phi_3\,.
\end{equation}

Hence we conclude, for example,
\begin{equation}
  LE_1(L)-\frac{(n-1)}{2n}LE_2(L)=2\pi(n^2-1)(n-2)f_3\,\alpha_\ms^4(1/L)+\dots\,,
\end{equation}
where $\alpha_\ms=\gbar_\ms^2/(2\pi)$.

In the next subsection we test this prediction
for $n=3$ and $n=4$ for which independent computations of the 
spectrum of excited states exist.
The first computations for O(3) were done by Balog and Heged\"us
\cite{Balog:2009ze} using the TBA 
equations with the knowledge of the exact Zamolodchikov
S-matrix \cite{Zamolodchikov:1977nu}.
Later Gromov, Kazakov and Vieira \cite{Gromov:2008gj} 
computed the excited states for O(4) from Hirota dynamics; 
in earlier papers Balog and Heged\"us \cite{Balog:2003yr,Balog:2005yz} 
computed the mass gap $E_1$ for O($2r$) but not $E_2$.   

\subsection{Running coupling functions}
\label{running}

For the perturbative analysis of their data, Balog and Heged\"us
\cite{Balog:2003yr}
introduced a function $\gbar_\J^2(L)$ of the box size $L$ 
\footnote{analogous constructions can be made for functions
  of lengths in infinite volume} through
\begin{equation}\label{gJ}
  \frac{1}{\gbar_\J^2(L)}+\frac{b_1}{b_0}\ln(b_0\gbar_\J^2(L))
  =-b_0\ln(\Lambda_{\FV}L)\,, 
\end{equation}
where $b_0\,,b_1$ are the universal first perturbative coefficients of the
$\beta-$function: 
\begin{equation}
  b_0=\frac{(n-2)}{2\pi}\,,\quad\quad
  b_1=\frac{(n-2)}{(2\pi)^2}\,,
\end{equation}
and $\Lambda_\FV$ is the $\Lambda-$parameter of 
the LWW finite volume coupling in \eqref{FVcoupling}. 

For $\Lambda_{\FV}L$ small 
there are usually two solutions of \eqref{gJ}
for $\gbar_\J^2(L)$, one large and one small
\footnote{The solution of the equation
$\frac{1}{\alpha}+\ln(\alpha)=-\ln(X)$ is 
$\alpha=-\frac{1}{W_{-1}(-X)}\,,$ where $W_{-1}(z)$
is the Lambert $W-$function \cite{Lambert},
first studied by Euler in 1783, and the index $-1$ refers 
to the real branch for which $\alpha\approx -1/\ln(X)\to +0$
when $X\to +0$.}.
We chose the solution which is small for $\Lambda_{\FV}L$ small, 
which has the property
\begin{equation}
\gbar_\J^2(L)=\gbar_{\FV}^2(L)+\order{\gbar_{\FV}^6(L)}\,,
\,\,\,\,\,\Lambda_{\FV}L\ll1\,.
\end{equation}
It is appropriate to call $\gbar_\J$ a ``running coupling function"
since it satisfies the equation
\begin{equation}
L\frac{\partial}{\partial L}\gbar_\J^2(L)
=b_0\gbar_\J^4(L)\left[1-\frac{b_1}{b_0}\gbar_\J^2(L)\right]^{-1}\,,
\end{equation}
i.e. like a perturbative running coupling with $\beta-$function
coefficients $b_r=b_0(b_1/b_0)^r\,,\,\,\,\,\forall r$.

Balog and Heged\"us consider $\gbar_J^2$ as a function of $z=ML$ where
$M$ is the infinite volume mass gap:
\begin{equation}
\frac{1}{\gbar_\J^2(z)}+\frac{b_1}{b_0}\ln(b_0\gbar_\J^2(z))
=-b_0\ln(z)+b_0\ln(M/\Lambda_{\FV})\,.
\end{equation}
The ratio $M/\Lambda_{\FV}$ is known from the work of
Hasenfratz and Niedermayer \cite{Hasenfratz:1990ab} 
\footnote{
$M/\Lambda_\msbar=(8/\mre)^\triangle/\Gamma(1+\triangle)$ and
$\Lambda_\FV/\Lambda_\ms=\exp\left\{-Z/2\right\}=
\Lambda_\ms/\Lambda_\msbar$.}
\begin{equation}
\frac{M}{\Lambda_\FV}=\frac{(8/\mre)^\triangle\mre^Z}{\Gamma(1+\triangle)}\,,
\quad\quad\triangle=\frac{1}{n-2}\,.
\end{equation}
Defining $\alpha_\J=\gbar_\J^2/(2\pi)$ the equation becomes \cite{Balog:2003yr}
\begin{equation} \label{aJ}
\frac{1}{\alpha_\J(z)}+\ln(\alpha_\J(z))=-(n-2)\ln(z)+J(n)\,,
\end{equation}
with
\begin{equation}
J(n)=(n-2)\left[Z-\ln(\Gamma(1+\triangle))\right]-\ln(n-2)-1+\ln(8)\,.
\end{equation}
In particular,
\begin{equation} \label{J34}
  J(3)=\ln(32\pi)-\gamma-1\,,\qquad J(4)=\ln(256\pi)-2\gamma-1\,.
\end{equation}

The LWW coupling has the following expansion in terms of $\gbar^2_\J$:
\begin{equation}
\gbar_{\FV}^2=\gbar^2_\J\left\{1+(n-2)\alpha_\J^2+j_3\alpha_\J^3+\dots\right\}
\,,
\end{equation}
with the coefficient $j_3$ given by
\begin{align}
j_3&=\frac18\left\{R_2(n-2)^3+\left(R_1-Z^2+4Z-\frac13\right)(n-2)^2\right.
\nonumber\\
&\left.+\left[R_0+6+6\zeta(3)+4Z\right](n-2)-2-6\zeta(3)
\phantom{\frac12}\right\}\,,
\end{align}
where the constants $R_i$ are given in \eqref{S0}--\eqref{S2}.
Inserting the numerical values \eqref{s1}--\eqref{s3} we obtain
\begin{equation}\label{j3n3}
j_3=1.195(4)\,,\,\,\,\,\,(n=3)\,,
\end{equation}
in agreement with the value quoted in eq.~(4.4) of \cite{Balog:2009ze}.

\subsection{\boldmath The Balog-Heged\"us results 
for the $I=1,2$ energies for O(3)} 

In table~\ref{tab:table1} we reproduce the data of 
Balog and Heged\"us \cite{Balog:2009ze};
($f_{3,\mathrm{est}}$ appearing in the last column is
defined in \eqref{f3est}).
Firstly we remark that by fitting the data for the mass gap $E_1$
one obtains a value of $j_3$ close to that 
given in \eqref{j3n3} deduced from Shin's analysis.

\begin{table}[ht]
  \centering
  \caption{O(3) energies for isospins $I=1,2$. }
  \label{tab:table1}
\vspace{0.5cm}
  \begin{tabular}{|l|l|l|l|l|l|}
    \hline
   \quad$z$ & $\quad\alpha_\J(z)$ &$\quad LE_1$ & $\quad LE_2$ 
   & $E_2/E_1$ & $f_{3,\mathrm{est}}$ \\
    \hline
    0.000001 & 0.05041270 & 0.31761386 & 0.95266912 & 2.9995 & 0.17707 \\
    0.000003 & 0.05354045 & 0.33744157 & 1.01210305 & 2.9993 & 0.17888 \\
    0.00001  & 0.05746199 & 0.36233411 & 1.08670379 & 2.9992 & 0.18159 \\
    0.00003  & 0.06159640 & 0.38862064 & 1.16546204 & 2.9990 & 0.18421 \\
    0.0001   & 0.06689778 & 0.42239721 & 1.26662427 & 2.9987 & 0.18785 \\
    0.0003   & 0.07263603 & 0.45905560 & 1.37636109 & 2.9982 & 0.19195 \\
    0.001    & 0.08023212 & 0.50775883 & 1.52204192 & 2.9976 & 0.19757 \\
    0.003    & 0.08877837 & 0.56282669 & 1.68656509 & 2.9966 & 0.20443 \\
    0.01     & 0.10066006 & 0.63994990 & 1.91651860 & 2.9948 & 0.21516 \\
    0.03     & 0.11489581 & 0.73342746 & 2.19419045 & 2.9917 & 0.23182 \\
    0.1      & 0.13647285 & 0.87841965 & 2.62093151 & 2.9837 & 0.27390 \\
    1.0      & 0.21988073 & 1.57045824 & 4.45605625 & 2.8374 & 0.72434 \\
    \hline
  \end{tabular}
\end{table}

Secondly, from the fifth column of table~\ref{tab:table1}, we see
that although the ratio $E_2/E_1$ is close to the ratio of the Casimir
eigenvalues, the data of Balog and Heged\"us establishes that 
the simple effective rotator model requires corrections. 

We have for the $I=2$ mass for $n=3$ the expansion
\begin{equation}
  LE_2(L)=6\pi\alpha_\J(z)\left\{1+\alpha_\J(z)^2
    +(j_3-8f_3)\alpha_\J(z)^3+\dots\right\}\,,\,\,\,\,(n=3)\,.
\end{equation}
With our value of $f_3$ we get a very small 3-loop coefficient
\begin{equation}
  j_3-8f_3=-0.007(4)\,,\,\,\,\,(n=3)\,,
\end{equation}
which is similar to the remark in \cite{Balog:2009ze}  
(before their eq.~(4.7)) that they numerically established that the 3-loop
coefficient was zero.
Although Shin's numbers (which have not yet been checked by an
independent computation) are consistent with the data, 
to indicate whether or not $j_3-8f_3$ is non-zero 
would require the values of the $s_i$ to a greater 
(at least a factor 10) accuracy than that obtained by Shin.

\begin{figure}[htb]
  \centering
  \includegraphics[width=0.9\textwidth]{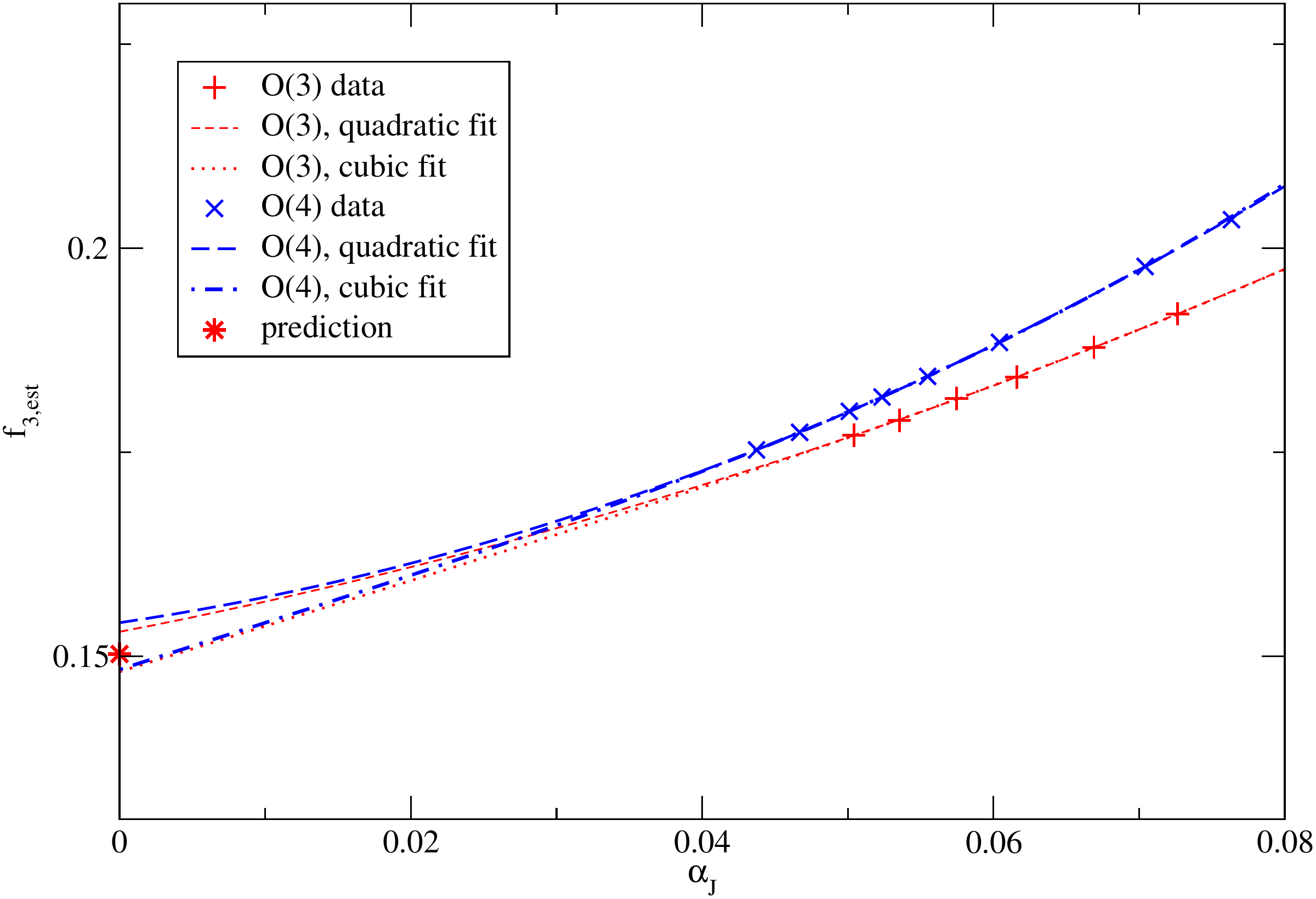}  
  \caption{\it Plot of the estimates for $f_3$ given in \eqref{f3est}
    for O(3) and O(4). 
    The curves are quadratic and cubic fits. 
    The star indicates the prediction in \eqref{f3} for $L\to 0$ limit.
  }
  \label{fig_f3est_n34}
\end{figure}

To see even more clearly the agreement of our analysis with
the data of Balog and Heged\"us,
in fig.~\ref{fig_f3est_n34} we plot estimates for $f_{3,\mathrm{est}}$ given by
\begin{equation}\label{f3est}
  f_{3,\mathrm{est}}=\frac{1}{(n^2-1)(n-2)2\pi\alpha_\J^4}
  \left[LE_1(L)-\frac{(n-1)}{2n}LE_2(L)\right]\,,
\end{equation}
for the case $n=3$ (and for $n=4\,,$ see next subsection).
Comparing the quadratic and cubic fits to the data in the range
$0.05 < \alpha_\J < 0.089$ (the first 8 points) 
one gets $f_3=0.151(2)$, which is in excellent
agreement with our prediction in \eqref{f3}.

\subsection{The O(4) case}

\begin{table}[ht]
  \centering
  \caption{O(4) energies for isospins $I=0,1,2$.
    The last column presents values of a 2-particle TBA parameter
    $\theta(z)=\theta_2(z)-\theta_1(z)$ discussed at the end of Appendix~C.4}
  \label{tab:table2}
  \vspace{0.5cm}
  \begin{tabular}{|l|l|l|l|l|}
  \hline
  \quad$z$ & $\quad L\mathcal{E}_0$ & $\quad L\mathcal{E}_1$ 
            & $\quad L\mathcal{E}_2$ & $\quad \theta$ \\
  \hline
    0.0005& $-1.358437256748$  &$-0.944622579962$ & $-0.255253656845$ 
&$3.399048890$\\
    0.001 & $-1.3434090793$  & $-0.9012815868$ & $-0.164826477882$ 
&$3.163405316$\\
    0.002 & $-1.325993755715$& $-0.8512533294$ & $-0.060589525136$ 
&$2.926136217$\\
    0.003 & $-1.31445222395$ & $-0.818220925$  & $\phantom{-}0.008145288724$ 
&$2.786480030$\\
    0.005 & $-1.29817998003$ & $-0.771823922$  & $\phantom{-}0.104558415733$ 
&$2.609523451$\\
    0.01  & $-1.27226343732$ & $-0.6983802787$ & $\phantom{-}0.256829671226$ 
&$2.367442907$\\
    0.03  & $-1.21827583735$ & $-0.547406298$  & $\phantom{-}0.568266040311$ 
&$1.9788498372$\\
    0.05  & $-1.185162243236$ & $-0.456340311$ & $\phantom{-}0.754882156720$ 
&$1.7962539024$\\
    0.1   & $-1.127336458694$ & $-0.300410888$ & $\phantom{-}1.071796663945$ 
&$1.5471797520$\\
    0.3   & $-0.9808905554557$ & $\phantom{-}0.07817485096$ 
& $\phantom{-}1.825340861386$ &$1.1532273953$\\
    0.5   & $-0.87023469365833$ & $\phantom{-}0.35536400274$ 
& $\phantom{-}2.363197150958$ &$0.9734482436$\\
    1.0   & $-0.64377457186332$ & $\phantom{-}0.938397059061$
& $\phantom{-}3.4676663603704$ &$0.7385633684$\\
    \hline                                                      
  \end{tabular}
\end{table}

In \cite{Balog:2003yr} the 1-particle data for $z\ge 0.001$ for the
O(4) model is sufficiently precise for our purposes, 
however, the 2-particle results are missing.
Gromov, Kazakov and Vieira \cite{Gromov:2008gj}
computed the 2-particle state energies, but for
$z<0.1$ we required higher precision than those presented in that reference.
For $z\ge 0.1$ we used the data obtained using the Mathematica program 
in ref.~\cite{Gromov:2008gj}
\footnote{We thank Nikolay Gromov for sending these data to us.}.

Afterwards we have rewritten the code, discretizing the
continuum formulation and using Fast Fourier Transform (FFT)
to calculate the convolutions appearing. Due to FFT the 
calculation of an iterative step becomes much faster and yields
more precise values.
In addition, approximating the 1d continuum functions by a finite
lattice of $N$ points, and using several $N$ values (typically 
$N=2^{10}\, \ldots 2^{14}$), one can get the continuum limit 
by determining numerically the coefficients of the corresponding
$1/N$ expansion.
Some details are given in Appendix~\ref{nu012}.

With this technique we could lower $z$, the size of periodic
box  down to $z=0.0005$\,.
For $z<0.0005$ it becomes more time consuming to produce reliable estimates
\footnote{Note that with decreasing $z$ the region
of attraction shrinks, and it becomes more and more difficult to start
the iterative method in the corresponding region.},
but the values of $\alpha_\J(z)$ (see table ~\ref{tab:table3})
are already quite small.

Our data for the TBA energies $\mathcal{E}_I$ for $I=0,1,2$
is presented in table ~\ref{tab:table2}.
The corresponding values for the excitation energies 
\begin{equation}
E_I(L)=\mathcal{E}_I(L)-\mathcal{E}_0(L)
\end{equation}
for isospins\footnote{In the sector with a given isospin $I$
only states with particle number $\nu \ge I$ contribute.
Among these the $\nu=I$ is the lightest, and the TBA equations
used describe just the $\nu=I=1$ and the $\nu=I=2$ states.},
$I=1,2$ together with $\alpha_\J$ are given in table~\ref{tab:table3}.

In that table we see that Casimir scaling
holds approximately for a large range of volumes.
Also we show estimates for $f_{3,\mathrm{est}}$ given by \eqref{f3est};
which are plotted in fig.~\ref{fig_f3est_n34} together with quadratic 
and cubic fits. 
In the range $0.0437 < \alpha_J < 0.075$ (7 points) one gets
$f_3=0.151(3)$.
Like in the O(3) case, this is consistent with our expected value 
$f_3=0.150257$  in \eqref{f3}.

\begin{table}[ht]
  \centering
  \caption{O(4) excitation energies for isospins $I=1,2$. }
  \label{tab:table3}
\vspace{0.5cm}
  \begin{tabular}{|l|l|l|l|l|l|}
    \hline
   \quad$z$ & $\quad\alpha_\J(z)$ &$\quad LE_1$ & $\quad LE_2$ 
   & $E_2/E_1$ & $f_{3,\mathrm{est}}$ \\
    \hline
    0.0005& $0.04373119$ & $0.41381468$ & $1.10318360$& $2.6659$ & $0.17527$ \\
    0.001 & $0.04669606$ & $0.44212749$ & $1.17858260$& $2.6657$ & $0.17743$ \\
    0.002 & $0.05010437$ & $0.47474043$ & $1.26540423$& $2.6655$ & $0.18001$ \\
    0.003 & $0.05234604$ & $0.49623130$ & $1.32259751$& $2.6653$ & $0.18176$ \\
    0.005 & $0.05548218$ & $0.52635606$ & $1.40273840$& $2.6650$ & $0.18429$ \\
    0.01  & $0.06041440$ & $0.57388316$ & $1.52909311$& $2.6645$ & $0.18846$ \\
    0.03  & $0.07041275$ & $0.67086954$ & $1.78654188$& $2.6630$ & $0.19776$ \\
    0.05  & $0.07633870$ & $0.72881932$ & $1.94004440$& $2.6619$ & $0.20349$ \\
    0.1   & $0.08627485$ & $0.82692557$ & $2.19913312$& $2.6594$ & $0.21551$ \\
    0.3   & $0.10919349$ & $1.05906541$ & $2.80623142$& $2.6497$ & $0.25110$ \\
    0.5   & $0.12497836$ & $1.22559870$ & $3.23343184$& $2.6382$ & $0.28403$ \\
    1.0   & $0.15648842$ & $1.58217163$ & $4.11144093$& $2.5986$ & $0.35723$ \\
    \hline                                  
  \end{tabular}
\end{table}

\section{\boldmath The case $d=3$}
\label{cased3}

From eqs.~(3.54)--(3.57) and (3.72)--(3.74) of \cite{Niedermayer:2016yll}
the expansion of the susceptibility with DR is given, 
already in \cite{Hasenfratz:1993vf},
\begin{equation}
  \chi = \frac{2\rho_s}{n} \left( 1 + \frac{1}{\rho_s L} \widetilde{R}_1
    + \frac{1}{(\rho_s L)^2} \widetilde{R}_2 + \ldots \right)\,,
\end{equation}
with for $D=3$:
\begin{equation}
  \widetilde{R}_1 = -\frac{(n-2)}{4\pi}\left(\gamma_2^{(3)}(\ell)-2\right)\,,
  \label{wtR1x} 
\end{equation}
and, (recall $\overline{W}=-L^{4-2D}\Psi$),
\begin{align}
  \widetilde{R}_2&= 2(n-2)\left\{ \Psi(\ell)
    +\frac{1}{32\pi^2}\left(\gamma_2^{(3)}(\ell)-2\right)
\left[2\alpha_1^{(3)}(\ell)-\gamma_2^{(3)}(\ell)-2-\frac{2}{\ell}\right]
\right.
\nonumber\\
&\left.+\frac{(n-2)}{32\pi^2\ell}\left(\gamma_3^{(3)}(\ell)+2\right)\right\}\,. 
  \label{wtR2x} 
\end{align}

We require the large $\ell$ behaviors of the shape functions
which we take from eqs.~(B.20) and (B.27) of \cite{Niedermayer:2017uyr}:
\begin{align}\label{alpha1elld3}
\alpha_1^{(3)}(\ell)&\simeq 
\alpha_{1/2}^{(2)}(1)+\frac13\pi\ell -2+\frac{1}{\ell}\,,
\\ \label{gamma2elld3}
\gamma_2^{(3)}(\ell)&\simeq
\alpha_{1/2}^{(2)}(1)-2+\frac23\pi\ell\,,  
\\ \label{gamma3elld3}
\gamma_3^{(3)}(\ell)&\simeq
\alpha_{3/2}^{(2)}(1)-\frac23+ \frac{4}{45}\pi^2 \ell^3\,.  
\end{align}

For the mass gap we have
from eqs.~(5.9)--(5.11) of \cite{Niedermayer:2016yll} (setting $\ellhat=1$)
\begin{equation}\label{E1_d3}
  E_1=\frac{(n-1)}{2\rho_s L^2}\left[1 
    +\frac{1}{\rho_s L}\widetilde{\triangle}^{(2)}
    +\frac{1}{(\rho_s L)^2}\widetilde{\triangle}^{(3)}+\dots\right]\,,
\end{equation}
with
\begin{align}
  \widetilde{\triangle}^{(2)}&=(n-2)LR(0)\,,
  \\ 
  \widetilde{\triangle}^{(3)}&=(n-2)L^2\left[
    -2L^{4-2D}\overline{\Psi}+\frac{3}{4L^{D-1}}\IDR_{10:D-1}+(n-3)R(0)^2\right]\,,
\end{align}
where we have set the coefficients of the four--derivative couplings
$l_1=l_2=0$.

From eq.~(3.16) of \cite{Niedermayer:2016ilf} for $D\sim3$:
\begin{equation}
  \IDR_{10:D-1}=-\frac{1}{2\pi L^{D-3}}
  \left[\frac{1}{D-3}-\frac12\alpha_1^{(2)}(1)+\frac12+\order{D-3} \right]\,.
\end{equation}
So for $d=3$ 
\begin{equation}\label{triangle3}
\widetilde{\triangle}^{(3)}=(n-2)\left[-2q_1
+\frac{3}{16\pi}\left\{\alpha_1^{(2)}(1)-1\right\}
+(n-3)L^2R(0)^2\right]\,,
\end{equation}
the $1/(D-3)$ singularities canceling as they should.

For the rotator we get
\begin{align}
\chi_{\mathrm{rot}}&=\frac{2\rho_s}{n}\left\{1-\frac{1}{\rho_s L}
\left[\frac{\ell}{6}(n-2)+\widetilde{\triangle}^{(2)}\right]
\right.
\nonumber\\
&\left.+\frac{1}{(\rho_s L)^2}\left[
\frac{\ell^2}{180}(n-2)(n-4)-\widetilde{\triangle}^{(3)}
+\widetilde{\triangle}^{(2)2}\right]+\ldots\right\}\,.
\end{align}

The expressions for $\chi$ and $\chi_{\mathrm{rot}}$
agree at 1-loop order up to terms vanishing exponentially with
$\ell$ since (see eq.~(5.10) of \cite{Niedermayer:2016ilf}) 
\begin{align}\label{Rzero}
L^{D-2}R(0)&=\frac{1}{4\pi}\int_0^\infty u^{-1/2}\left[S(u)^{D-1}-1\right]
\\
&=\frac{1}{4\pi}\left(\alpha_{1/2}^{(D-1)}(1)-\frac{2(D-1)}{D-2}\right)
\\ \label{LR0d3}
&=\frac{1}{4\pi}\left(\alpha_{1/2}^{(2)}(1)-4\right)\,\,\,\,\mathrm{for}
\,\,\,D=3\,.
\end{align}

Then one can check, using eqs.~\eqref{alpha1elld3}--\eqref{gamma3elld3}
and \eqref{Psielld3}, that at 2-loop $\chi$ and $\chi_{\mathrm{rot}}$ 
differ only by terms which vanish as $\ell\to\infty\,,$ and 
neglecting terms which vanish exponentially we obtain
\begin{equation}
\frac{\chi-\chi_{\mathrm{rot}}}{\chi}
= \frac{(n-2)(n+1)}{16\pi^2 \ell}
   \left(\alpha_{3/2}^{(2)}(1)+\frac43\right)(\rho_s L)^{-2}+\dots
\end{equation}

Suppose the spectrum is given by \eqref{ml}
and that $\Phi(L)$ has the expansion starting at order $(\rho_s L)^{-4}$:
\begin{equation}
\Phi(L)=\sum_{r=3}\Phi_r(\rho_s L)^{-r-1}\,,
\end{equation}
then from our considerations of the modified rotator
(cf eq.~(11.5) of \cite{Niedermayer:2017uyr}) 
\begin{equation}
\frac{\chi-\chi_{\mathrm{rot}}}{\chi}
= -\frac{4}{\ell}\Phi_3(n+1)(\rho_s L)^{-2}+\dots
\end{equation}
which determines 
\begin{equation}
\Phi_3=-\frac{1}{64\pi^2}\left[\alpha_{3/2}^{(2)}(1)+\frac43\right](n-2)\,,\,\,
\,\,\,\,\,(d=3)\,.
\end{equation}

Nyfeler and Wiese \cite{Nyfeler_thesis} have
investigated the histogram of the uniform
magnetization for the spin 1/2 Heisenberg model on the
honeycomb lattice in the cylindrical regime 
(see eqs.~(3.9)--(3.12) and fig.~4.3).
Having previously determined the low-energy constants
$\rho_s$ and $c$ for this model in the cubical regime,
they took the NLO formula for $\Theta$ and the standard rotator spectrum,
to compare the cylindrical regime. The agreement seemed to be very
good, but their statistical errors are too large
to see signs of our predicted $\mathrm{Casimir}^2$ term  
(or even the NNLO term in $\Theta$).
We are not aware of any other measurements of the spectrum
of the $d=3$ O$(n)$ models.

\section{Acknowledgements}
We would like to thank Janos Balog, Nikolay Gromov
and Uwe-Jens Wiese for useful correspondence.

\begin{appendix}
  
\section{\boldmath Spectrum of O(3) rotator in $d=1$ 
at finite lattice spacing}
\label{spectrumdeq1}

The aim here is to illustrate that the energy levels at a given order 
of the lattice spacing are polynomials of $C_I=I(I+1)$, not only of $I$.

The eigenvalues of the transfer matrix are given by
\begin{equation} \label{lI_def}
  \lambda_I(\epsilon) = \frac{1}{2\epsilon} \int_{-1}^{1}\mrd z\, 
\mre^{-(1-z)/(2\epsilon)}P_I(z)\,,
\end{equation}
where $P_I(z)$ are the Legendre polynomials.
These are polynomials in $1-z$ with coefficients which are 
polynomials of $C_I$ of order $I$:
\begin{equation}\label{PIexp}
  P_I(z) = \sum_{n=0}^I (1-z)^n \frac{(-1)^n}{2^n (n!)^2}
  \prod_{k=0}^{n-1} (C_I-C_k)\,.
\end{equation}

In the continuum limit one should take $\epsilon\to 0$.
Neglecting non-perturbative, exponentially small terms 
$\propto \mre^{-1/\epsilon}\,,$  one obtains the expansion 
\footnote{Note that the summand vanishes for $n>I$ hence the upper limit 
of the sum could be extended to infinity.}
\begin{equation} \label{lIexp}
  \lambda_I(\epsilon) \simeq \sum_{n=0}^{I}\frac{(-1)^n \epsilon^n}{n!} 
\prod_{k=0}^{n-1} (C_I-C_k)\,.
\end{equation}
For the energies one obtains
\begin{multline}
  E_I(\epsilon) = -\frac{1}{\epsilon} \ln\left(\lambda_I(\epsilon)\right)
  = C_I (1 + \epsilon) - \frac13 C_I(C_I-6) \epsilon^2   \\
  +\frac52 C_I^2\left( 1-\frac13 C_I + \frac{1}{60} C_I^2\right) \epsilon^3 
 + \order{\epsilon^4}\,.
\end{multline}
Here the coefficients of $\epsilon^n$ are indeed polynomials in $C_I$.

\section{The sunset diagram for $D\sim2,3,4$ 
  for large $\ell$}
\label{sunset_ell}

The sunset diagram for the susceptibility is
(cf. \cite{Niedermayer:2016ilf} (4.1) and (4.36))
\begin{equation}
  \begin{split}
    \Psi(\ell,\ellhat) & = -L^{2D-4}\Wov
    L^{2D-4} \int_{V_D} \mrd x\, \ddot{G}(x) G^2(x)\\
    &= - \frac{1}{48\pi^2(D-4)}\left( 10\, \ddot{g}(0;\ell,\ellhat)
      - \frac{1}{\mcVD} \right) -\frac{1}{16\pi^2}\mcWov(\ell)
    + \order{D-4}\,,
  \end{split}
  \label{Psi_d4}
\end{equation}
here we have reinstated $\ellhat$, but since we are working with
$\ellhat=1$ we have $\mcVD=\ell\ellhat^{D-4}=\ell$.

For the analogous diagram in the infinite strip one had
(cf. eqs.~(5.11) and (5.61) in \cite{Niedermayer:2016ilf} )
\begin{equation}
    \overline{\Psi}(\ellhat) = -L^{2D-4}W=
L^{2D-4} \int_{V_\infty} \mrd x
    \,\ddot{R}(x) R^2(x)
    = - \frac{1}{48\pi^2(D-4)} 10\, \ddot{R}(0;\ellhat) - c_w  
    + \order{D-4}\,.
  \label{ovPsi_d4}
\end{equation}
Further, for $\ell\gg 1$ one has up to exponentially small corrections
(cf. eq.~(5.20) in \cite{Niedermayer:2016ilf})
\begin{equation}
  \ddot{g}(0;\ell,\ellhat)- \frac{1}{\mcVD} \simeq \ddot{R}(0;\ellhat)\,.
\end{equation}

In Appendix~(B.4) of ref.~\cite{Niedermayer:2017uyr} we derived the relation 
\begin{equation}
  \frac{1}{16\pi^2} \mcWov(\ell) - \frac{1}{180} \ell^2
  +\frac{1}{12} \beta_1^{(3)}(1) \left(\ell+\frac92\right)
  +\frac{1}{16\pi^2}
    \left( \frac32 \alpha_0^{(3)}(1)-1\right)\frac{1}{\ell}
    \simeq c_w\,.
  \label{R2T2rel}
\end{equation}

\subsection{Case $D\sim2$}
\label{sunset_d2_ell}

For $D\sim2$
(setting $\ellhat=1$)
\begin{equation}
    \Psi(\ell) = \frac{1}{8\pi^2}\left[\frac{1}{(D-2)^2}
    +\frac{r_\Psi(\ell)}{(D-2)}
    +\overline{w}(\ell)\right]+ \order{D-2}\,,
  \label{Psi_d2}
\end{equation}
with 
\begin{align}
r_\Psi(\ell)&\equiv -\alpha_1^{(2)}(\ell)-\frac12+\frac{1}{\ell}
\\
&\simeq -\alpha_{1/2}^{(1)}(1)+\frac32-\frac{\pi}{3}\ell\,.
\end{align}

For the analogous diagram in the infinite strip 
\begin{equation}
    \overline{\Psi}=\frac{1}{8\pi^2}\left[\frac{1}{(D-2)^2}
    +\frac{p_0}{(D-2)}+p_1\right] + \order{D-2}\,.
  \label{ovPsi_d2}
\end{equation}

Using similar methods to that used to derive \eqref{R2T2rel}
we obtain for $D\sim2$:
\begin{equation}
  \begin{split}
    \Delta\Psi &\equiv \Psi(\ell)-\overline{\Psi}
    \\ 
    & \simeq  \frac{\ell^2}{180}
     +\frac{1}{16\pi^2}\left[
   \frac{3\pi}{2}\left(\alpha_1^{(1)}(1)+1\right)
    -\frac{3}{2\ell}\left(\alpha_{3/2}^{(1)}(1)+\frac13\right) \right.
   \\
   & \qquad \left.
   -\frac{2\pi\ell}{3(D-2)}
   +\frac{\pi\ell}{3}\left(\alpha_{1/2}^{(1)}(1)-2\right)
   \right]\,.
  \label{DeltaPsix}
  \end{split}
\end{equation}
Comparing this with \eqref{Psi_d2} and \eqref{ovPsi_d2} requires
\begin{equation}
  p_0=-\alpha_{1/2}^{(1)}(1)+\frac32= Z-\frac12\,,
\end{equation}
and
\begin{equation}
  \overline{w}(\ell)-p_1
  \simeq \frac{2\pi^2\ell^2}{45}-\frac{\pi\ell}{6}Z
  + \frac{\pi^2}{4}-\frac{3}{4\pi\ell}\zeta(3) \,.
  \label{R2T2reld2}
\end{equation}

\subsection{Case $D\sim3$}
\label{sunset_d3_ell}

For $d=3$ (setting $\ellhat=1$) 
\footnote{There is a printing error in eq.~(4.44) 
of ref.~\cite{Niedermayer:2016ilf}; for 
$d=3$, $\Psi=-L^2\overline{W}$.}
\begin{equation}
    \Psi(\ell) =-\frac{1}{16\pi^2}\overline{\mathcal{W}}(\ell)\,, 
  \label{Psi_d3}
\end{equation}
with values given in eq.~(4.45) of ref.~\cite{Niedermayer:2016ilf}
\begin{equation}  \label{mcWov_d3}
  \overline{\mathcal{W}}(\ell)=
  \begin{cases}
    2.12506105522294\,,\qquad & \text{for } \ell=1\,,\\
    1.90198910547056\,,\qquad & \text{for } \ell=2\,.
  \end{cases}
\end{equation}

For $D\sim3$ $\overline{\Psi}$ has a singularity:
\begin{equation} \label{OverlinePsielld3}
  \overline{\Psi}=-\frac{3}{16\pi(D-3)}+q_1+ \order{D-3}\,.
\end{equation}
Again using similar methods to that used to obtain \eqref{R2T2rel},
we can derive the expansion of $\Delta\Psi$ for $D\sim3$ for large $\ell\,,$
and find that the expansion of $\Psi(\ell)$ large $\ell$ is given by:
\begin{equation} \label{Psielld3}
  \begin{split}
    \Psi(\ell)& \simeq  -\frac{\ell^2}{180}+q_1
    -\frac{3}{32\pi}\left[\alpha_1^{(2)}(1)-1\right]
    \\
    &-\frac{\ell}{48\pi}\left[\alpha_{1/2}^{(2)}(1)-4\right]
    +\frac{3}{32\pi^2\ell}\left[\alpha_{3/2}^{(2)}(1)+\frac43\right]\,.
  \end{split}
\end{equation}

For the purposes of this paper we do not need the value of the constant
$q_1$ appearing in eqs.~\eqref{OverlinePsielld3} and \eqref{Psielld3}.
However we can easily obtain its numerical value from
the lattice computation of the perturbative coefficients
of the moment of inertia given in eq.~(7.7) of \cite{Niedermayer:2010mx}: 
\begin{align}\label{theta1_d3}
\theta_1&=\phantom{-}0.310373220693\,(n-2)\,,
\\ \label{theta2_d3}
\theta_2&=-0.000430499941\,(n-2)\,.
\end{align}
These coefficients are simply related to the coefficients
in \eqref{E1_d3} through
\begin{equation}
\theta_1=-\widetilde{\triangle}^{(2)}\,, \qquad
\theta_2=-\widetilde{\triangle}^{(3)}+\widetilde{\triangle}^{(2)2}\,.
\end{equation}
Using eq.~\eqref{triangle3} and the numerical values
\begin{align} \label{alpha2x1half}
\alpha^{(2)}_{1/2}&=0.0997350799980441171545246633952\,,
\\ \label{alpha2x1}
\alpha^{(2)}_1&=0.100879698927913998245389274826\,,
\end{align}
we obtain
\begin{equation}\label{theta2_d3x}
\theta_2=\left[2q_1+0.149993826254627007174930112530\right](n-2)\,.
\end{equation}
Combining this with eq.~\eqref{theta2_d3} we get
\begin{equation} \label{q1}
q_1=-0.075212163098\,.
\end{equation}

As a check, using eqs.~\eqref{alpha2x1half} and \eqref{alpha2x1} and
\begin{equation}
\label{alpha2x3half}
\alpha^{(2)}_{3/2}{\color{blue}(1)}=0.104412211554310173598670103056\,,
\end{equation}
we obtain, by comparing \eqref{Psielld3} with \eqref{mcWov_d3} at $\ell=2\,,$
the estimate $q_1\approx-0.07521075$ which agrees with \eqref{q1}
to nearly 5 digits \footnote{For $d=4$ the analogous
estimate for $c_w$ at $\ell=2$ was correct to 7 digits.}. 

\section{\boldmath The equations for the lowest energy
  state in a finite volume for the vacuum, 1- and 
  2-particle sectors}
\label{nu012}

We consider here the SU(2)$\times$SU(2)
principal chiral model which is equivalent to the O(4) non-linear sigma model.

Al. and A.~Zamolodchikov \cite{Zamolodchikov:1977nu}
proposed a self consistent exact S-matrix 
for the principal chiral models, which is built from the factors
\begin{equation} \label{S0x}
  S_0(x)= -i \frac{\Gamma(1+ix/2)\Gamma(1/2-ix/2)}%
  {\Gamma(1-ix/2)\Gamma(1/2+ix/2)}\,.
\end{equation}

In the equations for $I=0,1,2$ discussed below the following kernels occur. 
\begin{align}
  \Sp(x) &= S_0(x+i/2) \,, \label{Sp}
  \\
  K_0(x) &=  \frac{1}{2\pi i} \frac{\mrd}{\mrd x} \left(
    \ln\left(S_0(x)^2 \right)\right) \notag
  \\
  &= \frac{1}{2\pi}\left[  \Psi(1+ix/2) +\Psi(1-ix/2) 
      - \Psi(1/2+ix/2) - \Psi(1/2-ix/2)\right]\,,   \label{K0x}
  \\
  \Km(x) &= K_0(x-i/2) \,,  \label{Kmx}
  \\
  \Kpp(x) &= K_0(x+i) \,.  \label{Kppx}
\end{align}
In the relations above it is assumed that $x$ is real.
Unlike the others, $\Kpp(x)$ has a pole term $-i/(\pi x)$.
Further, it satisfies the relation
\begin{equation} \label{KppK0}
  \Kpp(x) 
  = -\left( K_0(x) - \frac{1}{\pi(1-i x)} + \frac{i}{\pi x}\right) \,.
\end{equation}
Below we shall consider separately the cases of the ground states 
in the sectors $I=0,1,2$. 

The equations to be solved are clearly presented in
ref.~\cite{Balog:2003yr}; our discussion, however is based
on the notations of ref.~\cite{Gromov:2008gj}.

\subsection{The iterative calculation of the vacuum energy}
The finite-volume vacuum energy is defined through a single function 
$A(x)$, which is obtained by an iterative solution.

The initial function is chosen for convenience as
\begin{equation} \label{A00x}
  A^{(0)}(x) = A_{00}(x) \equiv -\exp\left( -z \cosh(\pi x) \right)\,.
\end{equation}
The iterative step with the input function $A^{(k)}(x)$
calculates the output function\footnote{We denote the output of
the iteration with $\hat{A}$ since it could be different
from the input of the next iteration.}
$\hat{A}^{(k)}(x)$ through the steps
$A^{(k)}(x) \to r(x) \to f(x) \to \hat{A}^{(k)}(x)$.
The auxiliary functions\footnote{For intermediate quantities like $r(x)$,
$f(x)$ we don't write the iteration number $k$, except for clarity
when they appear in the final answer.}
are defined below. The reason for using $\hat{A}^{(k)}$ for the output,
is that the input of the next iteration, $A^{(k+1)}(x)$,
will be modified to improve the convergence of the iterations.

After introducing 
\begin{equation} \label{rkx_n0}
  r(x) = \ln\left( \frac{A^{(k)}(x)-1}{|A^{(k)}(x)|-1}\right) \,,
\end{equation}
and $r_c(x)=r(x)^*$, the function $f(x)$ is given by 
\begin{equation} \label{fxa_n0}
  f(x) = \left( K_0*r - \Kpp*r_c\right)(x)
  -r_c(x)\,.
\end{equation}
Here '$*$' denotes convolution. 
Due to the singularity of $\Kpp(x)$ at $x=0$ a Principal 
Value integral appears in \eqref{fxa_n0} instead of an ordinary convolution,
which is more difficult to deal with numerically.

One can circumvent this problem by introducing the function
\begin{equation} \label{Phix}
  \Phi(x)= \frac{i}{\pi}\left(\frac{1}{x+i} + \frac{1}{x-i\epsilon}\right)\,,
\end{equation}
which besides of the PV integral includes also the extra $-r_c(x)$
in \eqref{fxa_n0}.
One obtains then a simpler relation with two convolutions only,
\begin{equation} \label{fxb_n0}
  f(x) = \left( K_0*(r+r_c)\right)(x)
      + \left( \Phi*r_c\right)(x)\,.
\end{equation}
Finally, the output function of the iteration step is
\begin{equation} \label{Akp1x}
  \hat{A}^{(k)}(x) = A_{00}(x) \exp\left( f^{(k)}(x) \right) \,.
\end{equation}

The vacuum energy after $k$-th iteration is given by
\begin{equation} \label{E0k_n0}
  \mathcal{E}_0^{(k)} 
  = - M \int_{-\infty}^\infty \mrd x\, \RE (r^{(k)}(x))\cosh(\pi x)\,.
\end{equation}

To improve convergence one can use the well known trick
which takes the input for the next iteration a linear combination 
\begin{equation} \label{imp_n0}
  A^{(k+1)}(x) = \alpha \hat{A}^{(k)}(x) + (1-\alpha) A^{(k)}(x) \,,
\end{equation}
with a parameter $0<\alpha\le 1$.
Note that the choice of the parameter $\alpha$ does not influence
the fixed point (provided that the iteration converges),
however, it affects the properties of the iteration step, 
like the speed of convergence or whether the initial function $A^{(0)}(x)$
is in the domain of attraction of the iterative process. 
One expects that increasing $\alpha$ increases the speed of convergence 
to the fixed point but decreases the convergence region.
The latter could be crucial for small $L$ (i.e. small $z$)
since decreasing $L$ also shrinks the domain of attraction. 

For later use, note that the FT of $\Phi(x)$ is quite simple,
\begin{equation} \label{tPhip}
  \tilde{\Phi}(p)=-2 \left(\Theta(p) + \mre^{p} \Theta(-p)\right)\,,
\end{equation}
where $\Theta(p)$ is the Heaviside step function.
The function $\tilde{\Phi}(p)$ is continuous, but its first derivative
has a finite jump at $p=0$.

\subsection{Using discrete FT}
\label{FT}

The discrete FT in Maple is defined as
\begin{equation}
  g_F[k] = \FT(g)[k] 
  =\frac{1}{\sqrt{N}} \sum_{j=0}^{N-1} \mre^{-i \,2\pi k j/N} g[j]\,,
\end{equation}
and its inverse
\begin{equation}
  g[j] = \IFT(g_F)[j] 
  =\frac{1}{\sqrt{N}} \sum_{k=0}^{N-1} \mre^{i \,2\pi k j/N} g_F[k]\,.
\end{equation}
Here
\begin{equation} \label{xp}
  \begin{aligned}
    & x_j=j \mrd x\,, \quad (j=0,\ldots,N-1)\,, \qquad
      p_k=k \frac{\mrd p}{2\pi}\,, \quad (k=0,\ldots,N-1) \\
    & \mrd x = \frac{Q}{\sqrt{N}}\,, \quad 
     \frac{\mrd p}{2\pi}=\frac{1}{Q \sqrt{N}}\,,\qquad  
      \mrd x \, \frac{\mrd p}{2\pi} = \frac{1}{N}\,, \\ 
    & \exp\left(2\pi \, i \frac{k j}{N}\right) = \exp(i p_k x_j)\,,
     \qquad g(x_j) = g[j] \,, \qquad \tilde{g}(p_k) \sim Q \, g_F[k] \,.
  \end{aligned}
\end{equation}

It is assumed that $g(x)$ is a periodic function with period 
$2X= N \mrd x$. Since the functions appearing here are decreasing 
fast for $|x|\to\infty$, for sufficiently large $N$ one can consider 
the function to be periodic, $g(-X) = g(X) = 0$.
To represent both the positive and negative $x$ values in an array 
with indices $1 \le j \le N$, we choose $N$ points with coordinates
$[x_j, j=1,\ldots,N]= [0,1,2,\ldots,N/2-1,-N/2,\ldots,-2,-1]{\mrd x}$.

For $N\to\infty$ the resolutions $1/{\mrd x}$ and $1/{\mrd p}$
both in the original $x$-space (which is here related to rapidity)
and in $p$-space are proportional to $\sqrt{N}$.
The (rapidity) cutoff is given by $X=N {\mrd x}/2 \propto \sqrt{N}$.
Therefore the $N$ behavior given in \eqref{xp} for constant $Q$
obeys both requirements: for $N\to\infty$ one
approaches the continuum limit and the infinite cutoff limit.
  
At this point $Q$ is independent on $N$ but otherwise arbitrary.
Its value can be chosen to represent $A(x)$ and $\tilde{A}(p)$
by their corresponding discretizations equally well.
The number of points in the interval $\Sigma_x$, the width of
the function $|A(x)|$, is $\Sigma_x/\mrd x \approx \sqrt{N}\Sigma_x/Q$. 
Similarly, for $\tilde{A}(p)$ the number of points in its relevant region
in Fourier space is $\Sigma_p/\mrd p \approx \sqrt{N}\Sigma_p Q /(2\pi)$.
The value of $Q$ is expected to be optimal when these two numbers
are roughly equal, i.e. $Q^2 \approx 2\pi \Sigma_x/\Sigma_p$. 

Further for the convolution one has
\begin{equation} \label{conv} 
  (g*h)(x) = \int \mrd y\, g(x-y)\, h(y) 
  = \int \frac{\mrd p}{2\pi} \, \tilde{g}(p)\, \tilde{h}(p) \mre^{ipx} 
  \sim Q \, \IFT(g_F h_F)(x)\,.
\end{equation}

Rewriting \eqref{fxb_n0} by performing a discrete Fourier transformation 
one gets
\begin{equation} \label{fAb}
  \begin{aligned}
    f(x) &= Q \, \IFT\left(K_{0F}(p) ( r_{F}(p) + r_{cF}(p)) \right)(x)
    \\
    &-2\, \IFT\left( \left[\Theta(p)+\mre^p \Theta(-p)\right]r_{cF}(p)
    \right)(x)\,.
  \end{aligned}
\end{equation}
(The factor $Q$ appears only in the first term!)

\subsection{\boldmath $N$-dependence of the sums 
  appearing in FT and the Euler-Bernoulli relation}

Assuming that the function $g(x)$ is smooth for $x\ne 0$,
and it has a jump in the first derivative, according to the 
Euler-Bernoulli relation one has
\begin{equation}
  \begin{aligned}
  & \int_{-\infty}^{\infty} \mrd x\, g(x) - a \sum_{n=-\infty}^{\infty}g(a n)
  \\  
  &=
  - B_2 \frac{\delta g'(0)}{2!} a^2 
  - B_4 \frac{\delta g'''(0)}{4!} a^4
  - B_6 \frac{\delta g^{(5)}(0)}{6!} a^6
  - \cdots
  - B_{2k}\frac{\delta g^{(2k-1)}(0)}{(2n)!} a^{2n}
  \\
  &= -\frac{1}{12}\delta g'(0) a^2
  + \frac{1}{720}\delta g^{(3)}(0)  a^4
    -\frac{1}{30240}\delta g^{(5)}(0) a^6 + \ldots
  \end{aligned}
\end{equation}
where $\delta g^{(n)}(0)=g^{(n)}(+0)-g^{(n)}(-0)$ is the jump
of the corresponding derivative at $x=0$. Here $a$ is the lattice spacing
representing $\mrd x$ or $\mrd p$.

In our case $a^2\propto 1/N$; hence under the assumption
mentioned above, the corresponding sums have a $1/N$ expansion
with integer powers. 
One can calculate a quantity for several $N$ values and determine 
from these the expansion coefficients
\footnote{Allowing in the fit odd powers of $1/N^{1/2}$ it turns out 
indeed that the coefficient of $1/N^{3/2}$ is suppressed by a factor 
of $\sim 10^{-5}$.}. The corresponding fits produce a very stable
continuum extrapolation, often to $\sim 12$ digits. 

It seems that it is better to use the naive non-improved sum, 
(appearing anyhow in the FT) whose 
$1/N$ behavior is known rather than an improved sum which 
has a more complicated (or even unknown) type of approach to 
the continuum limit. In \cite{Gromov:2008gj} the authors took
$\alpha=1/2$.

\subsection{\boldmath The energy of the 1-particle sector}
\label{E_1p}

For the 1-particle sector
the input of an iteration step consists of a function
$A^{(k)}(x)$ and a number $\theta^{(k)}$.
Within an iteration one has the chain $[A^{(k)}(x), \theta^{(k)}] 
\to r(x) \to f(x) \to
[\hat{A}^{(k)}(x), \hat{\theta}^{(k)}]\,,$ (cf. \cite{Gromov:2008gj}).
The $A^{(k)}(x) \to r(x) \to f(x)$ part is given by the same
expressions we had in \eqref{rkx_n0} - \eqref{fxb_n0}.

In \cite{Gromov:2008gj} the authors choose a general $\theta^{(0)}$
as a starting point of iterations.
However, one can verify that $\theta^{(k)}\to 0$ for $k\to\infty$.
The physical reason is that $\theta$ is related to the rapidity
of the particle \cite{Balog:2003yr}.\footnote{In the expression of the 
energy\cite{Gromov:2008gj} for general $\theta$ the first term  
in \eqref{E1_n1} is replaced by $M \cosh(\pi\theta)$. 
Note that the second term will also depend on $\theta$ through
$r^{(k)}(x)=r^{(k)}(x;\theta)$.}
  
The state in the $I=1$ sector with the smallest energy 
is the 1-particle state with zero momentum, corresponding to $\theta=0$. 
Starting the iteration with $\theta^{(0)}=0$ all the subsequent $\theta^{(k)}$
values remain automatically zero.
For this reason, we set $\theta=0$, simplifying the iteration steps
to $A^{(k)}(x) \to \hat{A}^{(k)}(x)$.

The output function after one iteration (for $\theta=0$) is
written in the form
\begin{equation}
  \hat{A}^{(k)}(x)=A_{00}(x)\left[\Sp(x)\right]^2 \exp(f^{(k)}(x))\,,
\end{equation} 
(cf. \eqref{Sp}).
The iteration $A^{(k)}(x)\to \hat{A}^{(k)}(x)$ has the same form
as for the vacuum, given by \eqref{rkx_n0}--\eqref{Akp1x}.
(Note that due to the factor $[\Sp(x)]^2$ the agreement is 
only formal.)

Introducing the averaging in this case, one obtains for the next
input function
\begin{equation}
  A^{(k+1)}(x) = \alpha \hat{A}^{(k)}(x)   + (1-\alpha) A^{(k)}(x)\,.
\end{equation}

The energy of the ground state in the 1-particle sector for $\theta=0$
after the $k$-th iteration is given by
\begin{equation} \label{E1_n1}
  \mathcal{E}_1^{(k)} = M
  -M \int\mrd x\,\RE\!\left(r^{(k)}(x)\right)\cosh(\pi x) \,.
\end{equation}

\subsection{\boldmath The ground state energy of the 
  2-particle sector}
\label{E_2p}

For the case of the 2-particle sector,
the input of an iteration step consists of a function $A^{(k)}(x)$
and two numbers, $\theta_1^{(k)}$ and $\theta_2^{(k)}$.
In this case again it is enough to restrict the iteration
to the zero total momentum, i.e. $\theta_1 + \theta_2=0$. 
(The iteration would drive the system anyhow to satisfy
this condition at the FP.) 

In the rest frame one has $\theta_2=-\theta_1=\theta/2$,
where $\theta$
is related to the rapidity difference.
Because of the symmetry $\theta \to -\theta$ we can restrict
the discussion to $\theta > 0$.

In the case of zero total momentum the input of the iteration
is $[A^{(k)}(x),\theta^{(k)}]$ and the output is 
$[\hat{A}^{(k)}(x),\hat{\theta}^{(k)}]$.

Consider the equation 
\begin{equation} \label{eq_q}
   z \sinh\left(\frac12 \pi\theta\right) + \IM\left[
    \ln\left(S_0^2(\theta)\right) + 2 \phi\left(\frac12\theta\right)\right]
  = 0\,,
\end{equation}
where we introduced the function (depending implicitly on the input variables)
\begin{equation} \label{phi_n2}
  \phi(x) = (\Km * r)(x) \,.
\end{equation}

The output value of $\theta$ is given by the positive solution
of \eqref{eq_q},
\begin{equation} \label{hth_n2}
  \hat{\theta}^{(k)} = \theta > 0 \,.
\end{equation}

The output function $\hat{A}^{(k)}(x)$ after the $k$-th iteration 
is given by
\begin{equation} \label{hA_n2}
  \hat{A}^{(k)}(x) = A_{00}(x)
  \left[\Sp\left(x+\frac12 \theta^{(k)} \right) 
    \Sp\left(x-\frac12 \theta^{(k)} \right)\right]^2
  \exp(f^{(k)}(x)) \,,
\end{equation}
where the procedure to obtain $f^{(k)}(x)$
is formally the same as for the $\nu=0,1$ cases. 
Here one can introduce two averaging parameters $\alpha_A$ 
and $\alpha_\theta$:
\begin{equation}
  A^{(k+1)}(x) = \alpha_A\, \hat{A}^{(k)}(x) + (1-\alpha_A) A^{(k)}(x) \,.
\end{equation}
and
\begin{equation}
  \theta^{(k+1)} = \alpha_\theta\, \hat{\theta}^{(k)} 
   + (1-\alpha_\theta) \theta^{(k)} \,.
\end{equation}
The presence of these two parameters plays an important role
in the stability of the iteration and for the speed of convergence.
For example, choosing $\alpha_\theta \ll 1$ the value of  
$\theta$ barely changes. In our O(4) simulations at small $z$
we found that it is much easier to find a good starting point, 
(one in the domain of attraction of the fixed
point) when one chooses $\alpha_\theta < \alpha_A$.
For our last point, $z=LM=0.0005$ we took $\alpha_A=0.4$, 
$\alpha_\theta=0.14$.

The energy of the ground state in the 2-particle sector 
after the $k$-th iteration is
\begin{equation} \label{E2_n2}
  \mathcal{E}_2^{(k)}=
  2 M \cosh\left(\frac12 \pi \theta^{(k)}\right)
  -M\int\mrd x\,\RE\!\left(r^{(k)}(x)\right)\cosh(\pi x) \,.
\end{equation}

For a given $z$ and $N$ we let the iteration run until
the result stabilized (apart from fluctuations due to finite precision)
yielding the FP value 
  $\theta^{(\infty)}(N;z)= \lim_{k\to\infty}\theta^{(k)}(N;z)$
for several $N$ values. (We used $N=2^{10}\,,\ldots\,, 2^{14}$).
With these one can calculate the first few coefficients
of the $1/N$ expansion, $\theta^{(\infty)}(N;z)=\theta(z)+t_1(z)/N +
t_2(z)/N^2 + \ldots$.
This gives $\theta=\theta(z)=\theta_2(z)-\theta_1(z)$
appearing in \eqref{eq_q} and tabulated in Table~\ref{tab:table2}.

From the numerical analysis described above we found that $\theta(z)$ 
has a logarithmic dependence for small $z$,
$\theta(z)\approx -\ln(z)/\pi + a$, where $a\simeq 1.0$.

Using the running coupling $\gbar_\J^2(z)$ (cf.~\eqref{gJ}, \eqref{aJ})
one obtains a good fit (with an error smaller than $10^{-3}$),
\begin{equation} \label{Th_gsq}
  \theta(z) \approx \frac{1}{\gbar_\J^2(z)} + A + B\, \gbar_\J^2(z) \,,
\end{equation}
where $A=-0.1754$, $B=-0.2385$.

As mentioned earlier, the domain of attraction seems to shrink
with decreasing $z$, therefore it is important to choose
a proper starting value $\theta^{(0)}$ for the iteration.
The value obtained by \eqref{Th_gsq} can be used for $\theta^{(0)}$.
Note, however, that we did not investigate systematically 
questions related to the domain of attraction.

\end{appendix}

\end{document}